\documentclass[11pt,twoside]{article}
\usepackage{asp2004}
\usepackage{psfig}
\usepackage{epsf}
\usepackage{graphics}
\usepackage{lscape}
\def\lesssim{\mathrel{\hbox{\rlap{\hbox{\lower4pt\hbox{$\sim$}}}\hbox{$<$}}}}

\def\edcomment#1{\iffalse\marginpar{\raggedright\sl#1\/}\else\relax\fi}
\reversemarginpar

\begin{document}
\title{Supernova Rates in Galaxy Clusters }
\author{Dan Maoz}
\affil{School of Physics and Astronomy, Tel-Aviv University, Tel-Aviv 
69978, Israel; maoz@wise.tau.ac.il}

\begin{abstract}
Measurements of SN rates in different environments
and redshifts can shed light on the nature of SN-Ia
progenitors, star formation history, and chemical
enrichment history. I summarize some recent 
work by our group in this area, and 
discuss the implications. 
The current evidence favors production of most 
of the iron in the ICM (and perhaps everywhere) by
core-collapse SNe, rather than SNe-Ia. These SNe may
have been produced by the first, top-heavy-IMF,
generation of stars that reionized the Universe.  
Improved rate measurements can sharpen the picture, and
I describe our recent efforts in this direction.
\end{abstract}
\thispagestyle{plain}

Learning 
what are the progenitors of the different types of SNe, what are their rates,
and what are their distributions in space and in cosmic time, 
are essential steps
toward understanding SN physics, cosmic metal enrichment, 
 and galaxy formation
(e.g., Kobayashi et al. 2000). 
Core-collapse
SNe explode promptly ($\lesssim 10$~Myr) after the formation of a
stellar population, and their rate traces the
star-formation history (SFH). 
 In contrast, SN-Ia explosions should occur only after 
 WD formation and binary evolution,  with a ``delay'' of order 0.1--10~Gyr.
The SN-Ia rate vs. cosmic time will therefore be a convolution of 
the SFH with a ``delay function,'' the SN-Ia rate following a brief
burst of star formation. 
Each of the SN-Ia progenitor scenarios 
predicts a different delay function (e.g., 
Madau et al. 1998; 
Yungelson \& Livio 2000), which in turn dictates the SN 
rate vs. redshift. 

SN rate measurements in rich galaxy clusters at redshifts
$0<z<1$ are particularly relevant for constraining the SFH
in different environments, the progenitors of the
different SN types, and the contribution of each type to cosmic metal
 enrichment.
 Clusters are useful places to study enrichment because their deep
 potentials make them ``closed boxes'' from which little matter can escape.
The intra-cluster medium (ICM), 
which contains about 90\% of the baryons in clusters, 
has a high metallicity.
The iron abundance, in particular, is the one most easily and robustly 
measured, because iron has the strongest X-ray emission line in clusters,
and because
the theoretical conversion from data to abundance is simple (the line is 
dominated by emission from He-like ions). It is widely agreed that massive
clusters have a ``canonical'' iron abundance of $0.3 Z_{\odot}$, with little
dependence on cluster mass (e.g., Lin et al. 2003) or redshift, 
out to $z\approx 1$ (Tozzi et al. 2003; Hashimoto et al. 2004). 
Furthermore, the iron production of
the various SN types is observationally well constrained --- the radioactive
decay of  nickel to cobalt to iron drives much of the optical luminosity
of SNe, and thus empirical SN light curves yield direct estimates of
the ejected iron mass. It is easy to show that, 
including also the mass of iron
locked in the stellar component of a cluster, the total iron mass in clusters
is surprisingly large (e.g., Renzini et al. 1993; Maoz \& Gal-Yam 2004). 
If core-collapse SNe were the prime source of 
the iron in clusters, there is a factor of $\sim 5$
 excess of observed iron mass,
compared with expectations, given the observed present-day stellar mass.
\begin{figure}[fh]
\plotfiddle{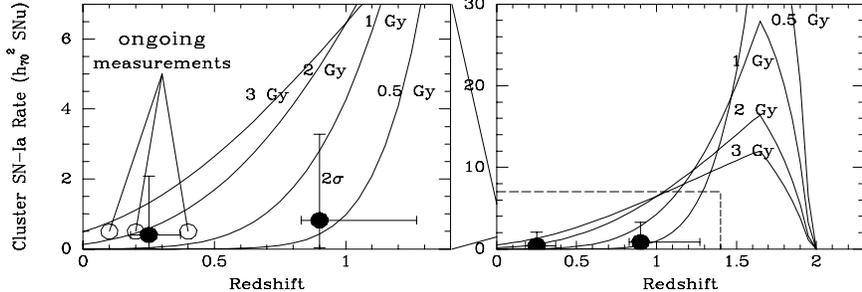}{1.05truein}{0}{74}{74}{-230}{-197}
\caption{
Cluster SN-Ia rates vs. redshift --- 
predictions and observations. Left
panel is a zoom-in on the low-redshift portion of the right 
panel. Rate curves follow the prescription of
Madau et al. (1998), for stellar formation redshift $z_f=2$
and SN-Ia delay times of 0.5, 1, 2, or 3~Gyr. The curves
are normalized to produce the observed iron-mass-to-light ratio
of rich clusters (Maoz \& Gal-Yam 2004). Filled circles are
the measurements of Gal-Yam et al. (2002) based on HST/WFPC2 data.
Empty circles illustrate the mean redshifts 
 of our three ongoing measurements at different redshifts,
plotted at arbitrary levels.
}
\label{snrates} 
\end{figure} 

A proposed solution to this puzzle is that clusters were enriched by a  
stellar population with a ``top-heavy'' IMF,
i.e., with relatively many high-mass stars, exploding as
iron-producing, core-collapse 
SNe --- perhaps the deaths of the first, very massive
stars, that reionized the Universe at $z\approx 10$ (e.g., Loewenstein 2001). 
Most of the
cluster stars seen today would then have formed
from the pre-enriched gas. Alternatively, most of the cluster iron may
have been produced by SNe-Ia from the normal stellar population. In principle,
one could discriminate between the two solutions using the observed 
elemental mix
in the ICM, relative to iron, compared to 
theoretical SN model yields. However, no consensus has emerged, due to the
observational and theoretical uncertainties in the analysis of 
non-iron elements (e.g., Buote 2002 vs. Lima-Neto et al. 2003).

 Measurement of SN rates vs. redshift (i.e., look-back time) 
offers a different, direct route of
investigating the source of the iron. The SN-Ia enrichment scenario 
requires a large total number of SNe-Ia, integrated over the cluster
lifetime. Since present SN-Ia rates are low, the SN-Ia rate must have been
much higher in the past. Indeed, for an assumed star-formation epoch in 
clusters ($z_f\approx 2$, based on the fundamental plane; van Dokkum \& Franx
2001) and a given SN-Ia delay function,
one can {\it predict} the cluster SN-Ia rate, SNR$_{Ia}(z)$,
required to produce the observed iron mass. Figure 1 shows an example of 
such a set of predictions. Measurement of SNR$_{Ia}(z)$ can not only
reveal the iron source directly (if it is SNe-Ia), but can also 
place observational
constraints on the SN-Ia delay function, and hence on the progenitor
scenario.

Gal-Yam, Maoz, \& Sharon (2002) carried out a
 first measurement of cluster SN-Ia rates
at high $z$ using HST archival data for eight rich clusters.
In Maoz \& Gal-Yam (2004), the measured rates 
at $z=0.25$ and $z=0.9$ were compared
to those predicted
by the SN-Ia iron production scenario. As seen in Fig. 1, the upper limit
on the SNR at $z\approx 1$ argues against models with long SN-Ia time delays,  
predicting SN rates ten times higher than observed. Thus, cluster 
iron production by 
SNe-Ia appears to be a viable option only if SNe-Ia have short ($\lesssim
 2$~Gyr)
time delays. Interestingly, comparison of the SNR$_{Ia}(z)$ in the {\bf field}
 to the cosmic star formation history suggests a {\bf long} delay time
(Pain et al. 2002; Tonry et al. 2003; Gal-Yam \& Maoz 2004; Strolger
 et al. 2004; Dahlen et al. 2004). 
The results therefore point to core-collapse SNe from an early,
top-heavy IMF, stellar generation as the source of cluster
 enrichment.
Alternatively, the existence of
two distinct SN-Ia progenitor populations 
-- an ``old'' one in clusters and a ``young''
one in the field -- is also an option
(Mannucci et al. 2004).  
 However, small-number statistics are the major limiting factor in 
current estimates of SNR$_{Ia}(z)$, and in constraints on
SN-Ia delay functions.   

To remedy this situation, we are engaged in obtaining
 accurate measurements of 
the SN-Ia rate in several samples of rich clusters. 
  Three different
observing programs, currently at various stages of execution, will
measure the rate in three different redshift ranges, as follows.\\
{\bf The Wise cluster SN survey; $0.08<z<0.2$ --}
In 1998--2001 we carried out a SN survey, using the Wise 1-m telescope,
on a sample of rich clusters at $0.08<z<0.2$. Among the results of
this survey were the first two clear examples of
intergalactic cluster SNe-Ia (Gal-Yam et al. 2002). Spectroscopic followup
of all candidates or their hosts has been completed at large
telescopes, mainly Keck.
We therefore now have a ``clean'' sample of seven
transients, each  spectroscopically confirmed as a cluster SN-Ia
(see Fig. 2), with
\begin{figure}[fh]
\epsfxsize=13.5cm
\epsfbox{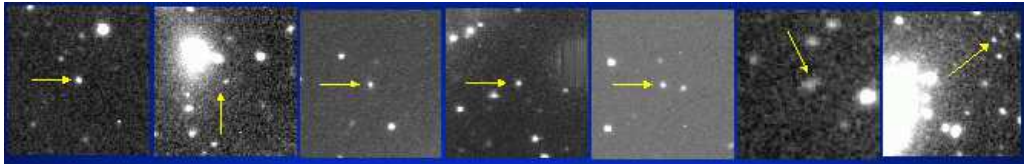}
\caption{Seven cluster SNe-Ia discovered at Wise and 
spectroscopically confirmed at Keck and other telescopes. From left
to right, SNe 1998eu, 1998fc, 1999cg, 1999ch, 1999ci, 1999ct, 2001al.
These SNe will be used to calculate the cluster SN-Ia rate at $0.08<z<0.2$.  
}
\label{snrates} 
\end{figure} 
all other candidates from the survey having been demonstrated to {\it not}
have been cluster SNe-Ia. To derive the SN-Ia cluster
rate in this redshift interval, we are
determining the detection efficiency. This
will allow computation of the control time for the survey, and hence the SN
rate (e.g., Gal-Yam et al. 2002).\\ 
{\bf The NOT cluster SN survey; $0.2<z<0.3$ --}
In the summer of 2004 we completed the imaging part of a cluster SN
survey in this redshift range, using the 2.6-m Nordic Optical
Telescope (NOT) in La Palma. A number of candidates have been
confirmed as cluster SNe-Ia using Keck spectroscopy (see Fig. 3 for an
example), 
\begin{figure}[fh]
\epsfxsize=11.5cm
\epsfbox{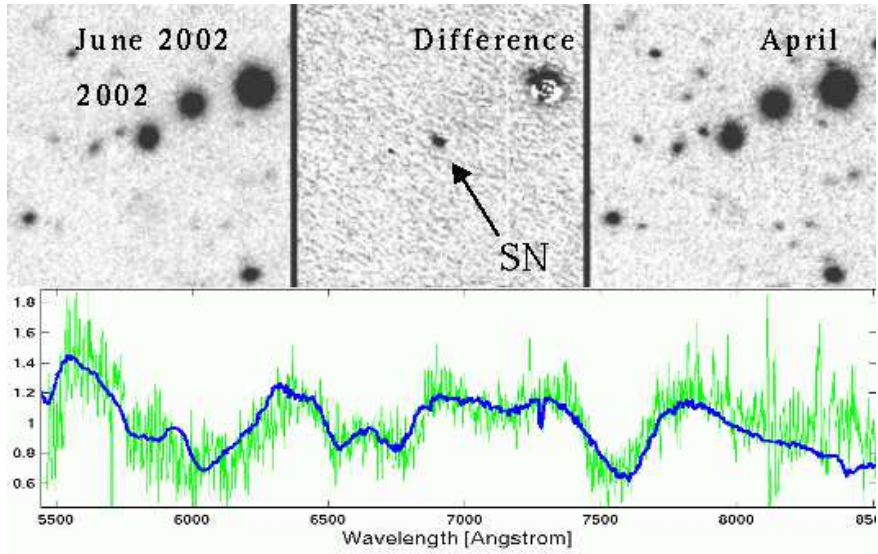}
\caption{Discovery images from the NOT of
  SN 2002lh, and a Keck spectrum 
showing it is a SN-Ia at the redshift of Abell 1961
  ($z=0.236$).
Solid curve is a spectrum of nearby SN-Ia 1999ee, suitably
  redshifted. This and other cluster SNe will be used to derive a
  cluster SN-Ia rate at $0.2<z<0.3$.
}
\label{snrates} 
\end{figure} 
and we continue 
the followup work on the less-promising transients --- these generally
have turned out to be active galactic nuclei (AGNs), 
variable stars, and SNe in the background
or foreground of a cluster. We will determine the detection
efficiencies for this survey as well, and derive the cluster SN-Ia
rate in this redshift interval.\\    
{\bf The Keck cluster SN survey; $0.3<z<0.5$} --
We have been granted Keck observing time (PI A. Gal-Yam) 
to measure the SN-Ia rate in a complete sample of X-ray-selected
$0.3<z<0.5$ clusters. The first
run, in October 2004, successfully obtained $B$ and $R$ reference
images for part of the sample. Subsequent runs
 will yield SNe. As in the nearer samples,
followup work and efficiency simulations will result in a SN-Ia
rate determination for this redshift bin.

\end{document}